\documentstyle[aps,prd,epsf]{revtex}

\newcommand{\be}{\begin{equation}}
\newcommand{\ee}{\end{equation}}
\newcommand{\ben}{\begin{eqnarray}
\displaystyle}
\newcommand{\een}{\end{eqnarray}}

\newcommand{\la}{{\lambda}}

\newcommand{\de}{{\delta}}

\newcommand{\cE}{{\cal E}}
\newcommand{\cP}{{\cal P}}

\newcommand{\cH}{{\cal H}}

\newcommand{\cJ}{{\cal J}}

\newcommand{\cO}{{\cal O}}
\newcommand{\cQ}{{\cal Q}}
\newcommand{\cR}{{\cal R}}
\newcommand{\cS}{{\cal S}}

\newcommand{\cL}{{\cal L}}

\newcommand{\na}{\nabla}

\newcommand{\LieN}{{\cal L}_{N^{i}}}
\newcommand{\Lie}{{\cal L}}

\newcommand{\tiB}{{\tilde B}}
\newcommand{\ticA}{{\tilde {\cal A}}}

\newcommand{\ep}{\epsilon}

\begin{document}

\title{Staticity Theorem for Higher Dimensional Generalized
Einstein-Maxwell System}

\author{Marek Rogatko}

\address{Institute of Physics \protect \\
Maria Curie-Sklodowska University \protect \\
20-031 Lublin, pl.Marii Curie-Sklodowskiej 1, Poland \protect \\
rogat@tytan.umcs.lublin.pl \protect \\
rogat@kft.umcs.lublin.pl}

\date{\today}

\maketitle
\begin{abstract}
We derive formulas for variations of mass, angular momentum and canonical energy in
Einstein $(n-2)$-gauge forms field theory by means of the ADM formalism.
Considering the initial data for the manifold with
an interior boundary which has the topology of $(n-2)$-sphere we obtained the generalized 
first law of black hole thermodynamics. Supposing that a black hole event horizon comprises 
a bifurcation Killing horizon with a bifurcate surface we find that the solution is static
in the exterior world, when the Killing timelike vector field is normal to the horizon 
and has vanishing {\it electric} or {\it magnetic} fields on static slices.
\end{abstract}


\baselineskip=18pt
\section{Introduction}
Nowadays there has been a significant resurgence of interest in gravity and black holes in more than
four dimensions. It stems from the attempts of building a consistent quantum gravity theory
in the realm of M/string theory as well as in the range of TeV gravity, where
the large or infinite dimensions are taken into account. Especially mathematical aspects of
classification of n-dimensional black holes attract recently more attention.
As far as the
problem of classification of non-singular black hole solutions in four-dimensions was concerned
Israel \cite{isr},
M\"uller zum Hagen {\it et al.} \cite{mil73} and Robinson \cite{rob77},
presented first proofs.
The most complete results were provided in Refs.\cite{bun87,ru,ma1,he1,he93}.
The classification of both static vacuum black hole solutions as well as
the Einstein-Maxwell (EM) black holes
was finished 
in \cite{chr99a,chr99b}.
\par
The problem of the uniqueness black hole theorem for stationary axisymmetric
spacetime turned out to be more complicated. It was elaborated
in Refs.\cite{stat}, but the complete proof was provided by
Mazur \cite{maz} and Bunting \cite{bun}
(see for a review of the uniqueness of black hole
solutions story see \cite{book} and references therein).
\par
Attempts of building a consistent quantum gravity theory triggered the researches 
concerning the mathematical aspects of the low-energy string theory
black holes.
The uniqueness of the black hole solutions in dilaton gravity was  proved in 
works
\cite{dil,mar01}, while the uniqueness of the static
dilaton $U(1)^2$
black holes being the solution of $N = 4, d = 4$ supergravity
was provided in \cite{rog99}. The extension of the uniqueness proof to the 
case of static dilaton black holes with 
$U(1)^{N}$  gauge fields was established in Ref.\cite{rog02}.
\par
On the other hand,
$n$-dimensional black hole uniqueness theorem, both in vacuum and
charged cases was given in Refs.\cite{gib03,gib02a,gib02b,kod04}. The case
of nonlinear self-gravitating $\sigma$-model in higher dimensions was treated in
\cite{rog02a}.
The complete 
classification of $n$-dimensional
charged black holes having both degenerate and non-degenerate
components of event horizon was provided in Ref.\cite{rog03}.\\
In Ref.\cite{mye86} it was pointed out that a black hole being the source of both 
magnetic and electric components of $2$-form $F_{\mu \nu}$ was a striking coincidence.
Hence, in order to treat
this problem in $n$-dimensional gravity one should consider
both {\it electric} and {\it magnetic} components of 
$(n-2)$-gauge form $F_{\mu_{1} \dots \mu_{n-2}}$. 
In Ref.\cite{rog04} the proof of the uniqueness
of static higher dimensional {\it electrically} and {\it magnetically} charged
black hole containing an asymptotically flat hypersurface with compact interior
and non-degenerate components of the event horizon was given.
\par
Proving the uniqueness theorem for stationary 
$n$-dimensional black holes is much more complicated.
It turned out that generalization of Kerr metric to arbitrary $n$-dimensions proposed by
Myers-Perry \cite{mye86} is not unique. The counterexample showing that a five-dimensional
rotating black hole ring solution with the same angular momentum and mass
but the horizon of which was homeomorphic to $S^{2} \times S^{1}$ was presented in
\cite{emp02} (see also Ref.\cite{emp04}). In Ref.\cite{mor04} it was shown that 
Myers-Perry solution is the unique black hole in five-dimensions in the class of 
spherical topology
and three commuting Killing vectors \cite{mor04}, while in \cite{rog04a} the problem of 
a stationary nonlinear self-gravitating $\sigma$-model in five-dimensional
spacetime was considered. It was proved that when we assume that the horizon had the
topology of $S^3$ then, the Myers-Perry vacuum Kerr solution
is the only one maximally extended, stationary, axisymmetric flat solution
having the regular rotating event horizon with constant mapping.
\par
The uniqueness theorem for black holes is closely related to the problem of staticity for
non-rotating black holes and circularity for rotating ones. For the first time the problem of
staticity was tackled by Lichnerowicz \cite{lic}. The next extension to the vacuum spacetime was
attributed to Hawking \cite{haw}, while the extension taking into account electromagnetic fields was
provided by Carter \cite{car}. But only recently 
the complete proof of staticity
theorem \cite{sud92,sud93} by means of the ADM formalism was given. In the case of the low-energy string theory
the problem of staticity was studied in Refs.\cite{rog97,rog98}.
\par
In our paper we shall study the problem of staticity theorem in the Einstein $(n - 2)$-gauge forms
$F_{(n-2)}$ theory. Sec.II will be devoted to the canonical formalism of the underlying theory.
In Sec.III we tackle the problem of canonical energy and angular momentum and derive the first
law of thermodynamics for black holes with $(n - 2)$-gauge forms
$F_{(n-2)}$ fields. 
Our derivation of the first law of black hole thermodynamics relies on the assumption
that the event horizon is a Killing bifurcation $(n - 2)$-dimensional sphere.
Then, we find the conditions for staticity for non-rotating black holes in
$n$-dimensions.
\par
In what follows the Greek indices will range from 0 to $n$. They denote 
tensors on an $n$-dimensional manifold, while the Latin ones run from 1 to $n$ and
denote tensors on a spacelike hypersurface $\Sigma$. The adequate covariant derivatives are signed 
respectively as $\na_{\alpha}$ and $\na_{i}$.

\section{Higher dimensional generalized Einstein-Maxwell system}
In this section we shall examine the generalized Maxwell $(n-2)$-gauge
form $F_{\mu_{1} \dots \mu_{n-2}}$ in $n$-dimensional spacetime
described by the following
action:
\be
I = \int d^n x \sqrt{-{g}} \bigg[ {}^{(n)}R - 
F_{(n-2)}^2
\bigg],
\label{act}
\ee
where ${g}_{\mu \nu}$ is $n$-dimensional metric tensor,
$F_{(n-2)} = dA_{(n-3)} $ is $(n-2)$-gauge form field.
The canonical formalism divides the metric into spatial and temporal parts, as follows:
\be
ds^2 = - N^2 dt^2 + h_{ab} \bigg( dx^{a} + N^{a}dt \bigg) \bigg( dx^{b} + N^{b}dt \bigg)
\ee
where general covariance implies the great arbitrariness in the choice of lapse and shift functions
$N^{\mu}~(N,~N^{a})$.
\par
A point in the phase space for the underlying theory is related to the specification of the fields
$(h_{ab},~ \pi_{ab},~ A_{j_{1} \dots j_{n-3}},~ E_{j_{1} \dots j_{n-3}})$ on
$(n-1)$-dimensional hypersurface $\Sigma$. The field momenta are found in the usual way by varying the
Lagrangian with respect to $\na_{0} h_{ab},~ \na_{0} A_{j_{1} \dots j_{n-3}}$, where 
$\na_{0}$ denotes the derivative with respect to the time coordinate.
Thus, the momentum canonically conjugates to a Riemannian metric is
\be
\pi^{ab} = \sqrt{h} \big( K^{ab} - h^{ab} K \big),
\ee
where $K_{ab}$ is the extrinsic curvature of the hypersurface ${\Sigma}$.
Similarly,
the momentum canonically conjugates to $(n-2)$-gauge form field $F_{\mu_{1} \dots \mu_{n-2}}$
is defined as
\be
\pi_{j_{1} \dots j_{n-3}}^{(F)} = {\de {\cal L} \over \de (\na_{0} A^{j_{1} \dots j_{n-3}})}
= 2 (n -2) E_{j_{1} \dots j_{n-3}}.
\ee
While the
{\it electric} field $E_{j_{1} \dots j_{n-3}}$ implies
\be
 E_{j_{1} \dots j_{n-3}} = \sqrt{h} F_{\alpha j_{1} \dots j_{n-3}}~n^{\alpha},
\ee
where $n^{\mu}$ is the unit normal timelike vector to the hypersurface ${\Sigma}$.
The Hamiltonian is defined by the Legendre transform 
may be written as follows:
\ben
\cH &=& \pi^{ab}~\na_{0} h_{ab} + \pi_{(F)}^{j_{1} \dots j_{n-3}}~\na_{0} A_{j_{1} \dots j_{n-3}}
- \cL(R,~F_{(n-2)}) \\ \nonumber
&=& N^{\mu} C_{\mu} + \ticA_{0 j_{2} \dots j_{n-3}} \tiB^{j_{2} \dots j_{n-3}} + \cH_{div},
\een  
where for brevity of the notation
we have denoted by $\ticA_{j_{1} \dots j_{n-3}} = (n - 3)!~ A_{j_{1} \dots j_{n-3}}$. On the hand,
the total derivative part of the Hamiltonian 
$\cH_{div}$ is given by
\be
\cH_{div} = 2 (n - 3)(n - 2) \na_{j_{1}} \bigg(
E^{{j_{1} \dots j_{n-3}}}~\ticA_{0 j_{2} \dots j_{n-3}} \bigg) + 
2 \sqrt{h} \na_{i} \bigg( {N_{j} \pi^{ij} \over \sqrt{h}} \bigg).
\ee
The gauge field $\ticA_{0 j_{2} \dots j_{n-3}}$ has no associated with it kinetic terms.
Therefore one can consider it as a Lagrange multiplier corresponding to the generalized
{\it Gauss law} of the form as follows:
\be
0 = \tiB^{j_{2} \dots j_{n-3}} =
2 (n - 3)(n - 2) \na_{j_{1}} \bigg(
E^{{j_{1} \dots j_{n-3}}} \bigg).
\ee
In our paper we shall consider
the asymptotically flat initial data, i.e.,
in asymptotic region of hypersurface $\Sigma$ which is diffeomorphic to
${\bf R}^{n-1} - B$, where $B$ is compact, one has the following conditions to be satisfied:
\ben
h_{ab} &\approx&  \de_{ab} + \cO ({1\over r}), \\
\pi_{ab} &\approx& \cO ({1\over r^2}), \\
A_{j_{1} \dots j_{n-3}} &\approx& \cO ({1\over r}), \\
E_{j_{1} \dots j_{n-3}} &\approx& \cO ({1\over r}).
\een
At infinity we also assume the standard behaviour of the lapse and shift functions, i.e.,
$N \approx 1 + \cO ({1\over r})$ and $N^{a} \approx \cO ({1\over r})$.\\
On the hypersurface $\Sigma$ the initial data are restricted to the constraint manifold on which
at each point $x \in \Sigma$ the following quantities vanish
\ben
0 &=& C_{0} = \sqrt{h} \bigg[
- {}{}^{(n-1)}R + {1 \over h} \bigg( \pi_{ij} \pi^{ij} - {1 \over 2 }\pi^{2} \bigg)
\bigg]
+ {(n - 2) \over \sqrt{h}} E_{j_{1} \dots j_{n-3}}~E^{j_{1} \dots j_{n-3}} +
\sqrt{h}
F_{j_{1} \dots j_{n-2}}~F^{j_{1} \dots j_{n-2}}, \\ \nonumber
0 &=& C_{a} = 2 (n - 2) F_{a j_{1} \dots j_{n-3}}~E^{j_{1} \dots j_{n-3}} -
2 \sqrt{h}~ \na^{i} \bigg( {\pi_{ia} \over \sqrt{h}} \bigg), \\ \nonumber
0 &=& \tiB^{j_{2} \dots j_{n-3}} =
2 (n - 3)(n - 2) \na_{j_{1}} \bigg(
E^{{j_{1} \dots j_{n-3}}} \bigg),
\een
where $\na_{j}$ is the derivative operator on $\Sigma$, while ${}{}^{(n-1)}R$ denotes the scalar curvature
with respect to the metric $h_{ab}$ on the considered hypersurface.\\
The equations of motion for this theory can be formally derived from the {\it volume
integral contribution} $\cH_{V}$ to the Hamiltonian $\cH$ and 
subject to the {\it pure constraint} form as follows:
\be
\cH_{V} = \int_{\Sigma} d\Sigma \bigg(
N^{\mu} C_{\mu} + N^{\mu} \ticA_{\mu j_{2} \dots j_{n-3}} \tiB^{j_{2} \dots j_{n-3}} \bigg).
\label{vol}
\ee
One can verify that the
change caused by arbitrary infinitisemal variations 
$(\de h_{ab},~ \de \pi^{ab},~ \de \ticA^{j_{1} \dots j_{n-3}},~ \de E_{j_{1} \dots j_{n-3}})$
of compact support, after integration by parts leads us to the expression
\be
\de \cH_{V} = \int_{\Sigma} d\Sigma \bigg(
\cP^{ab}~ \de h_{ab} + \cQ_{ab}~ \de \pi^{ab} +
\cR^{j_{1} \dots j_{n-3}}~ \de \ticA^{j_{1} \dots j_{n-3}} + \cS^{j_{1} \dots j_{n-3}}~
\de E_{j_{1} \dots j_{n-3}} \bigg),
\ee
where $\cP^{ab}~,\cQ_{ab},~\cR^{j_{1} \dots j_{n-3}},~\cS^{j_{1} \dots j_{n-3}}$ are written as
\ben
\cP^{ab} &=& \sqrt{h}~N~a^{ab} + \sqrt{h}~ \bigg( h^{ab} \na^{j}\na_{j} N - \na^{a}\na^{b} N \bigg)
- \LieN \pi^{ab}, \\ \label{co1}
\cQ_{ab} &=& {N \over \sqrt{h}} \bigg( 2\pi_{ab} - \pi h_{ab} \bigg) + 2 \na_{a} N_{b}, \\ \label{co2}
\cR^{j_{1} \dots j_{n-3}} &=& - 2(n - 2) \bigg[
~ \na_{a} \bigg(
F^{a j_{1} \dots j_{n-3}} \bigg) +
~ \LieN E^{j_{1} \dots j_{n-3}}
\bigg], \\ \label{co3}
\cS^{j_{1} \dots j_{n-3}} &=&  {2 (n - 2) \over \sqrt{h}}~ N~ E^{j_{1} \dots j_{n-3}} +
2 (n - 2)~ \LieN \ticA^{j_{1} \dots j_{n-3}} + 2 (n - 2) (n - 3) \na^{j_{1}} \bigg(
N \ticA_{0}{}{}^{j_{2} \dots j_{n-3}} \bigg).
\label{co4}
\een
While $a^{ab}$ takes the form as
\ben
a^{ab} &=& {1 \over 2} F_{j_{1} \dots j_{n-2}}F^{j_{1} \dots j_{n-2}} - (n - 2)F^{a j_{2} \dots j_{n-2}}
F^{b}{}{}_{ j_{2} \dots j_{n-2}} \\ \nonumber
&-& {(n - 2) \over 2 \sqrt{h}} h^{ab}
E_{j_{1} \dots j_{n-3}} E^{j_{1} \dots j_{n-3}} + {(n - 2) (n - 3) \over \sqrt{h}}
E^{a j_{2} \dots j_{n-3}} E^{b}{}{}_{j_{2} \dots j_{n-3}} \\ \nonumber
&+& {1 \over h} \bigg[
2 \pi^{a}{}{}_{j} \pi^{bj} - \pi h^{ab} - {1 \over 2} h^{ab} \bigg(
\pi_{ij} \pi^{ij} - {1 \over 2} \pi^{2} \bigg) \bigg]
+ {}{}^{(n - 1)}R^{ab} - {1 \over 2} h^{ab}~{}{}^{(n - 1)}R^{ab}.
\een
In the above relations $\LieN$ denotes
the Lie derivative calculated on the hypersurface $\Sigma$.
The Lie derivative of $E^{j_{1} \dots j_{n-3}}$ is understood as the Lie derivative of the
adequate tensor density. 
\par
On using the Hamiltonians principle and evaluating variations of the compact support of $\Sigma$,
we finally reach to the evolutions equations which can be written as follows:
\be
{\dot{\pi}}^{ab} = - \cP^{ab},
\label{c1} 
\ee
\be{\dot{h}}_{ab} = \cQ_{ab}, 
\label{c2}
\ee
\be
{\dot{E}}^{j_{1} \dots j_{n-3}} = - \cR^{j_{1} \dots j_{n-3}}, 
\label{c3}
\ee
\be
{\dot{\ticA}}_{j_{1} \dots j_{n-3}} = \cS_{j_{1} \dots j_{n-3}}. 
\label{c4}
\ee
As was mentioned
in Ref.\cite{sud92} expression (\ref{vol}) depicts rather the {\it volume integral}
contribution to the Hamiltonian. The non-vanishing surface terms arise when we take into account
integration by parts. In order to get rid of these surface contribution terms one 
can add the surface terms
given by
\ben
\cH = \cH_{v} &+& \int_{S^{\infty}} dS_{j_{1}}
\bigg[ 
2(n - 2)~N^{a} \ticA_{a j_{2} \dots j_{n-3}} E^{j_{1} \dots j_{n-3}} +
2 (n - 2)(n - 3)~N~ \ticA_{0 j_{2} \dots j_{n-3}} E^{j_{1} \dots j_{n-3}} 
\bigg] \\ \nonumber
&+& \int_{S^{\infty}} dS^{a} 
\bigg[
N \bigg( \na^{b} h_{ab} - \na_{a} h_{b}{}{}^{b} \bigg) + {2 N^{b} \pi_{ab} \over \sqrt{h}}
\bigg].
\een
By the direct calculations it can be seen that not only for asymptotically flat perturbations of
a compact support of
the hypersurface $\Sigma$ but also for $N^{\mu}$ and $\ticA_{j_{1} \dots j_{n-3}}$ satisfying
the asymptotic conditions at infinity, we get
\be
\de \cH = 
\int_{\Sigma} d\Sigma \bigg(
\cP^{ab}~ \de h_{ab} + \cQ_{ab}~ \de \pi^{ab} +
\cR^{j_{1} \dots j_{n-3}}~ \de \ticA^{j_{1} \dots j_{n-3}} + \cS^{j_{1} \dots j_{n-3}}~
\de E_{j_{1} \dots j_{n-3}} \bigg).
\ee

\section{First law of black hole mechanics}
We can define the canonical energy as the Hamiltonian function corresponding to the case
when $N^{\mu}$ is an asymptotical translation at infinity. Thus, one has that 
$N \rightarrow 1,~ N^{a} \rightarrow 0$. We multiply Hamiltonian function by $1/2$
and reach to the expression
\be
\cE = \alpha~M + (n - 2)(n - 3) \int_{S^{\infty}} dS_{j_{1}} N~\ticA_{0 j_{2} \dots j_{n-3}}
E^{j_{1} \dots j_{n-3}},
\ee
where $M$ is the ADM mass defined as follows:
\be
\alpha M = {1 \over 2} \int_{S^{\infty}} dS^{a} \bigg[
N \big( \na^{b} h_{ab} - \na_{a} h_{j}{}{}^{j} \big) \bigg],
\ee
and $\alpha =  {n - 3 \over n - 2}$.
The remaining term is highly gauge dependent because of the arbitrary
choice of $\ticA_{0 j_{2} \dots j_{n-3}}$. It yields
\be
\cE_{F} = (n - 2)(n - 3) \int_{S^{\infty}} dS_{j_{1}} N~\ticA_{0 j_{2} \dots j_{n-3}}
E^{j_{1} \dots j_{n-3}}.
\ee
We shall call $\cE_{F}$ {\it canonical energy} of $(n - 2)$-gauge forms fields.\\
We define also 
the canonical angular momentum $\cJ_{(i)}$ 
on the constraint submanifold of the phase space as the Hamiltonian $\cH$
multiplied by the factor $1/2$,
when $N \rightarrow 0$ and the shift vector tends to the appropriate Killing vector fields
responsible for rotation in the adequate directions. Thus, it reduces to
\be
\cJ_{(i)}^{(\infty)} = - {1 \over 2} \int_{S^{\infty}} dS_{a} \bigg(
2 \phi_{(i)}{}{}^{b} \pi_{b}{}^{a} + 2(n - 2)(n - 3)~\phi_{(i)}{}{}^{m} \ticA_{m j_{2} \dots j_{n-3}}
E^{a j_{2} \dots j_{n-3}} \bigg).
\ee
If one considers the case of hypersurface $\Sigma$ having an asymptotic region and smooth
interior boundary $S$ and take into account the linear combinations of 
the translation and rotations at infinity, then we reach to the following expression:
\ben
2 \bigg( 
\de \cE &-& \sum\limits_{i = 1}^{n-1} \Omega_{(i)} \cJ^{(i) (\infty)} \bigg) \\ \nonumber
&=& \int_{\Sigma} d\Sigma \bigg(
\cP^{ab}~ \de h_{ab} + \cQ_{ab}~ \de \pi^{ab} +
\cR^{j_{1} \dots j_{n-3}}~ \de \ticA^{j_{1} \dots j_{n-3}} + \cS^{j_{1} \dots j_{n-3}}~
\de E_{j_{1} \dots j_{n-3}} \bigg) \\ \nonumber
&+&
\de (surface~ terms).
\een
As in Ref.\cite{sud92} one can take an asymptotically flat hypersurface $\Sigma$
which intersects the sphere $S$ of a stationary n-dimensional black hole. 
We assume also that $(n - 2)$-sphere $S$ is a bifurcation Killing horizon and
set that 
$N^{\mu} = \chi^{\mu} = t^{\mu} + \sum\limits_{i = 1}^{n-1} \Omega_{(i)} \phi^{\mu (i)}$,
where $\Omega_{(i)}$ describe angular velocities of the direction established
by $\phi^{\mu (i)}$. We also choose
$\ticA_{0 j_{2} \dots j_{n-3}}$ so that ${\dot A}_{j_{1}\dots j_{n-3}}
= {\dot E}_{j_{1}\dots j_{n-3}} = 0$. Using Eqs.(\ref{c1})-(\ref{c4})
one can draw a conclusion that the integral over $\Sigma$ vanishes while, the only one surface term
survives because of the fact that on sphere $S$ we have $N^{\mu} = 0$. The non-zero
term is equal to $2 \pi \kappa \de A$, where $\kappa$ is the surface gravity
constant on $S$, while $A$ is the area of the $(n - 2)$-dimensional sphere $S$. Thus we
reach to the following:\\
{\it Theorem:}\\
Let $(h_{ij},~ \pi^{ij},~\ticA_{j_{1} \dots j_{n-3}},~E^{j_{1} \dots j_{n-3}})$
be smooth asymptotically flat initial data for a stationary black hole
with $(n-2)$-gauge forms field
on a hypersurface $\Sigma$ with $(n-2)$-dimensional bifurcation sphere $S$.
If $(\de h_{ij},~\de \pi^{ij},~\de \ticA_{j_{1} \dots j_{n-3}},~\de E^{j_{1} \dots j_{n-3}})$
are arbitrary smooth asymptotically flat solutions of the linearized constraints on a
hypersurface $\Sigma$, then the following is fulfilled:
\be
\alpha~ \de M + \de \cE_{F} - \sum\limits_{i = 1}^{n-1} \Omega_{(i)} \cJ^{(i)(\infty)} = \kappa~ \de A.
\label{mass}
\ee
Taking into account (\ref{mass}) one can see that any stationary black hole
with a bifurcate Killing horizon is an extremum of mass $M$ at fixed
{\it canonical energy} of $(n - 2)$-gauge forms fields, canonical momentum and horizon area.\\
We get the extension of the first law of black hole mechanics which is true for 
arbitrary asymptotically flat perturbations of a stationary $n$-dimensional
black hole (in four-dimensions the similar result was obtained by Sudarsky and Wald \cite{sud92} in
Einstein Yang-Mills theory
and in the case of Einstein-Maxwell axion-dilaton black holes in Ref.\cite{rog97}).
Contrary to the first law of black holes mechanics derived in Ref.\cite{bar73}
valid for perturbations to a nearby stationary black hole.

\section{Staticity conditions}
Now we proceed to 
find the staticity theorem for non-rotating $n$-dimensional black holes
with $(n - 2)$-gauge field $F_{\mu_{1} \dots \mu_{n-2}}$. To begin with let us 
suppose that a stationary black hole is regular on and outside a Killing horizon
of a Killing vector field of the form
\be
\chi^{\mu} = t^{\mu} + \sum\limits_{i = 1}^{n-1} \Omega_{(i)} \phi^{\mu (i)},
\ee
is normal.
The mass of a black hole implies \cite{mye86} the following:
\be
M = - {1 \over \alpha} \int_{S} \ep_{j_{1} \dots j_{n-2}a b} \na^{a} t^{b}.
\ee
Furthermore, we define
the angular momentum of black hole associated with a rotational Killing vector
$\phi_{(i)}$ expressed as a covariant surface integral
\be
I_{(i) BH} = {1 \over 2} \int_{H} \ep_{j_{1} \dots j_{n-2}a b} \na^{a} \phi^{b}_{(i)}.
\ee
The same procedure as in Ref.\cite{bar73} leads us to the mass formula
\be
M = {2 \over \alpha} \int_{\Sigma} d \Sigma \bigg(
T_{\mu \nu} + {g_{\mu \nu} T \over 2 - n} \bigg) t^{\mu} n^{\nu}
+ {2 \over \alpha} \kappa A + {2 \over \alpha}  \sum\limits_{i = 1}^{n-1} \Omega_{(i)} I^{(i)}_{ BH}.
\label{bch}
\ee
Rewriting the latter expression (\ref{bch}) in 
terms of the considered matter energy momentum tensor yields
\ben
M &-& {2 \over \alpha} \kappa A - {2 \over \alpha}  \sum\limits_{i = 1}^{n-1} \Omega_{(i)} I^{(i)}_{ BH}
\\ \nonumber
&=& \int_{\Sigma} d \Sigma \bigg[
{(n - 2) \over \sqrt{h}} t^{m} F_{m j_{1} \dots j_{n-3}} E^{j_{1} \dots j_{n-3}}
+ {\la \over h}  E_{j_{1} \dots j_{n-3}} E^{j_{1} \dots j_{n-3}} +
\la \bigg( {n - 3 \over n - 2} \bigg) F_{j_{1} \dots j_{n-2}} F^{j_{1} \dots j_{n-2}}
\bigg],
\een
where we defined by $\la = - n_{\beta}t^{\beta}$.\\
Taking account of constraint equations and changing
the surface integral into a volume one 
we can deduce that $\cJ_{(i)}^{(\infty)}$ has the form as 
\be 
\cJ_{(i)}^{(\infty)} = - {1 \over 2} \int_{\Sigma} d \Sigma
\bigg( \pi^{ab}~\LieN h_{ab} + 2(n - 2)~\LieN \ticA_{j_{1} \dots j_{n-3}} E^{j_{1} \dots j_{n-3}}
\bigg) + \cJ_{(i) H},
\ee
where we define $\cJ_{(i) H}$ by the following expression:
\be
\cJ_{(i) H} = - {1 \over 2} \int_{S} dS_{a} 
\bigg[
2 \phi_{(i)}{}{}^{b} \pi_{b}{}^{a} + 2(n - 2)(n - 3)\phi_{(i)}{}{}^{m} \ticA_{m j_{2} \dots j_{n-3}}
E^{a j_{2} \dots j_{n-3}} 
\bigg].
\label{jj}
\ee
Using the fact that Killing vector fields $\phi_{\mu}^{(i)}$ are equal to their tangential
projection $\phi_{m}^{(i)}$
one can readily find
that $\cJ_{(i)}^{(\infty)} = \cJ^{(i)}_{H}$. The first term in relation (\ref{jj})
is equal to $I^{(i)}_{ BH}$. 
Then, from
(\ref{jj}) it follows immediately the result
\be
 \sum\limits_{i = 1}^{n-1} \Omega_{(i)} \bigg(
I^{(i)}_{ BH} - \cJ^{(i)~(\infty)} \bigg) =
(n - 2) \int_{\Sigma} d \Sigma \bigg[
E^{j_{1} \dots j_{n-3}}
\LieN \ticA_{j_{1} \dots j_{n-3}}
- t^{m} F_{m j_{1} \dots j_{n-3}} E^{j_{1} \dots j_{n-3}} \bigg].
\ee
By virtue of the above equation and the constraint relation (\ref{c4}) we find
the expression of the form
\ben \label{mmm}
M &-& {2 \over \alpha} \kappa A + {2 \over \alpha} \cE_{F} - {2 \over \alpha}\sum\limits_{i = 1}^{n-1} \Omega_{(i)}
\cJ^{(i)~(\infty)}  \\ \nonumber
&= &\int_{\Sigma} d \Sigma~ \bigg[
2 \la~ F_{j_{1} \dots j_{n-2}} F^{j_{1} \dots j_{n-2}}
- 2 {\la (n - 2) \over h} E_{j_{1} \dots j_{n-3}}E^{j_{1} \dots j_{n-3}}
\bigg].
\een
\par
From this stage on 
we shall restrict our attention to the case of
the maximal hypersurface, i.e., for which
$\pi_{a}{}{}^{a} = 0$. Having this in mind we consider the initial data induced
on hypersurface $\Sigma$ and choose the lapse and shift function coinciding
with Killing vector fields in the spacetime under considerations. It may be verified that contracting
Eq.(\ref{c1}) we get
\be
\na_{m} \na^{m} N = \rho~N,
\label{rr}
\ee
where $\rho$ is given by
\be
\rho = \bigg( {n - 3 \over n - 2} \bigg)~F_{j_{1} \dots j_{n-2}}F^{j_{1} \dots j_{n-2}}
+ {n \over 2 h (n - 2)} \pi_{ij} \pi^{ij} - {1 \over 2 h} \bigg[
(n - 1)(n - 5) + (3 - n) \bigg]~E_{j_{1} \dots j_{n-3}} E^{j_{1} \dots j_{n-3}}.
\ee
We remark that $\rho$ will be non-negative for $n \ge 4$. Consistently with this remark the maximum
principle can be applied to the relation (\ref{rr}) provided that solutions of it can be uniquely 
determined by their boundary value at $S$ and their asymptotic value at infinity.
\par
To proceed further, we use as the lapse function
$\la$ with the
boundary conditions $\la |_{S} = 0,~\la |_{\infty} =1$. Integrating Eq.(\ref{rr}) we obtain
{\it black hole mass formula} as
\be
M - {2 \over \alpha} \kappa A = {2 \over \alpha} \int_{\Sigma} d \Sigma~ \la~ \rho.
\label{bhm}
\ee
\par
Using the scaling transformation we can transform a solution of Einstein $(n-2)$-form gauge theory
into a new one with the following changes:
\ben
M &\rightarrow& \beta^{n-3} M, \\
\cE_{F} &\rightarrow& \beta^{n-3} \cE_{F}, \\
\Omega_{(i)} &\rightarrow& \beta^{-1} \Omega_{(i)}, \\
\cJ^{(i)~(\infty)} &\rightarrow& \beta^{n-2} \cJ^{(i)~(\infty)},\\ 
\kappa &\rightarrow& \beta^{-1} \kappa, \\
A &\rightarrow& \beta^{n-2} A,
\een
where $\beta$ is a constant. Inserting the linearized perturbation connected with the above 
scaling
transformation into equation (\ref{mass}) one finally left with the second mass formula of the form
\be
\alpha~ M - 2 \kappa A - 2~ \sum\limits_{i = 1}^{n-1} \Omega_{(i)} \cJ^{(i)~(\infty)}
+ \cE_{F} = 0.
\label{mass1}
\ee
Then, using (\ref{mmm}),~(\ref{bhm}),~(\ref{mass1}) one solves them for $\cE_{F}$ and
$\sum\limits_{i = 1}^{n-1} \Omega_{(i)} \cJ^{(i)~(\infty)}$. The results become as        
\be
\cE_{F} = \int_{\Sigma} d \Sigma \bigg[
4 \la \bigg( {n - 3 \over n - 2} \bigg)~ F_{j_{1} \dots j_{n-2}}F^{j_{1} \dots j_{n-2}} +
4 \la {n - 3 \over h}~E_{j_{1} \dots j_{n-3}} E^{j_{1} \dots j_{n-3}}
\bigg]
\ee
while the formula for the angular momenta can be written as
\ben \label{jjj}
\sum\limits_{i = 1}^{n-1} \Omega_{(i)} \cJ^{(i)~(\infty)} &=&
\int_{\Sigma} d \Sigma \bigg[
3 \la \bigg( {n - 3 \over n - 2} \bigg)~ F_{j_{1} \dots j_{n-2}}F^{j_{1} \dots j_{n-2}} +
{\la n \over 2 h (n - 2)}~ \pi_{ij} \pi^{ij} \\ \nonumber
&+&
{\la \over h} \bigg(
{3 (3 - n) - (n - 1)(n - 5) \over 2 (2 - n)}
 \bigg)~ E_{j_{1} \dots j_{n-3}} E^{j_{1} \dots j_{n-3}}
\bigg]
\een
In the case of four-dimensional spacetime the coefficient for $E_{(n-3)}^2$ in
Eq.(\ref{jjj}) is equal to zero. Thus we have the same result for $n = 4$ as was obtained in Ref.\cite{sud93}.
\par
In Ref.\cite{chr94} was pointed out that the exterior region of a black hole can be
foliated by maximal hypersurfaces with boundary $S$, asymptotically orthogonal to the timelike
Killing vector field $t_{\mu}$, when the strong energy condition for every timelike vector
is satisfied. As one can check this is the case in the considered theory. 
In the light what has been shown above, 
we can establish the following:\\
{\it Theorem}:\\
Let us consider an asymptotically flat solution to Einstein $(n-2)$-gauge theory possessing
a stationary Killing vector field and describing a stationary black hole comprising
a bifurcate Killing horizon. Suppose, moreover that 
$\sum\limits_{i = 1}^{n-1} \Omega_{(i)} \cJ^{(i)~(\infty)} = 0$,
then the solution is static and has vanishing {\it electric} $E^{j_{1} \dots j_{n-3}}$
or {\it magnetic} $F_{j_{1} \dots j_{n-2}}$ fields on static hypersurfaces.\\
\par
One can readily verify the above by applying Eq.(\ref{jjj})
to the maximal hypersurfaces. It will be noticed that
on the considered hypersurfaces $\Sigma_{t}$ one has the condition 
$\pi_{ij} = 0$. Let $N$ denotes the lapse function for the maximal
hypersurface and $n^{\mu}$ depicts unit normal to this hypersurface. We choose
$N^{\mu} = N n^{\mu}$ as the evolution vector field for these slices.
This is sufficient to establish that
\be
\Lie_{N^{\mu}} \pi^{ij} = {\dot \pi}^{ij} = 0.
\ee
From Eqs.(\ref{co1}) and (\ref{c2}) , since $\pi^{ab} = 0$ and $N^{a} = 0$, we obtain
that $\Lie_{N^{\mu}} h_{ab} = {\dot h}_{ab} = 0$.\\
We shall 
first consider the case when  $E^{j_{1} \dots j_{n-3}} = 0$. 
Consequently it yields the result as follows:
\be
\Lie_{N^{\mu}} E^{j_{1} \dots j_{n-3}} = {\dot E}^{j_{1} \dots j_{n-3}} = 0.
\ee
It can be verified that
considering Eqs.(\ref{co3}) and (\ref{c4})
and choosing $A_{0j_{2} \dots j_{n-3}} = 0$ one
gets that ${\dot \ticA}_{j_{1} \dots j_{n-3}} = 0$. 
By virtue of this we can conclude that the solution is static.
\par
Now we take into account the case when $F_{j_{1} \dots j_{n-2}} = 0$. To begin with
let us consider relation (\ref{co2}) from which because of the fact that $N^{a} = 0$,
one has that
$\Lie_{N^{\mu}} E^{j_{1} \dots j_{n-3}} = 0$.
Thus, we see that ${\dot E}^{j_{1} \dots j_{n-3}} = 0$. Now consider Eq.(\ref{c4}) and
Eq.(\ref{co4}) and choose $A_{0j_{2} \dots j_{n-3}} = 0$ as well as $E^{j_{1} \dots j_{n-3}} = 0$.
Then one can draw a conclusion that
${\dot \ticA}_{j_{1} \dots j_{n-3}} = 0$ and the solution is static.

\vspace{0.5cm}



\end{document}